# Nature of ferroelectric transitions in spin ice $Ho_2Ti_2O_7$ and $Dy_2Ti_2O_7$

**Pramod K. Yadav and Chandan Upadhyay**

School of Materials Science and Technology, Indian Institute of Technology (Banaras Hindu University), Varanasi 221005, India
Emails: pyadav.rs.mst13@iitbhu.ac.in, cupadhyay.mst@iitbhu.ac.in

*Abstract:* To investigate the possible origin and mechanism of ferroelectricity in polycrystalline spin ices $Ho_2Ti_2O_7$ and $Dy_2Ti_2O_7$ a detailed dielectric study has been performed. Experimental finding suggests that both materials have two prominent diffuse ferroelectric phase transitions around 90K and 36K. These transitions are distinctly generated by the lattice distortions at the oxygen sites as confirmed by triggered distortions and order of activation energy. Due to the incompatibility of the gyrotropic order with any phonon mode at the Brillouin zone center, observed diffuse ferroelectric phase transitions can have only an electronic origin. Through magnetic susceptibility and previously reported spin relaxation behavior it has been concluded that single ion anisotropy has thermal variation, due to which orientation of rare earth magnetic moment from isotropic non-Ising to Ising spin along local <111>axis takes place. This spin orientation distinctly distorting the both oxygen sites of the structure reflects in the form of diffuse ferroelectric phase transitions.

*Introduction:* In geometrically frustrated magnetic rare earth compounds, spin dynamics are mainly govern by competing mechanism originating from- conduction electrons[1], phonon mediation[2] or quantum mechanical phenomena[3,4]. Currently, the understanding of spin relaxation mechanism of quantum states in rare earth single ion magnets is a challenging task for their applications in quantum information processing[5,6]. In spin ices ($Ho_2Ti_2O_7$ and $Dy_2Ti_2O_7$), due to strong single ion anisotropy (SIA), each rare earth (RE) spin behaves as single ion magnets[7]. This anisotropy changes the RE magnetic moment in such a way that its maximum magnitude and direction lies along local <111> axis of $Fd\bar{3}m$ space group, termed as classical Ising spin[8–11]. The Hamiltonian for crystal field originating from SIA can be given as-

$$H_{cf} = -D\sum_i <S_i.n_i>^2 \qquad (1)$$

Where $n_i$ is a unit vector along the Ising <111> axis of the i$^{th}$ spin $S_i$ and D is the SIA coefficient[12]. At low temperatures, these locally constrained Ising spins are well-modeled by dipolar and exchange interactions which form local 2in-2out spin structure, similar to hydrogen ion arrangement in water ice[13,14], termed as spin ice state.

Recently, existence of ferroelectricity in these spin ices have attracted renewed research interest from the prospective of theoretical and experimental studies[15–17]. Experimental finding suggest that $Ho_2Ti_2O_7$ (HTO) and $Dy_2Ti_2O_7$ (DTO) spin ices have multiple ferroelectric transitions of different origin[15–18]. In polycrystalline HTO two ferroelectric transitions are observed at ~60K and ~23K[15], however in case of single crystal of HTO, only one transition, originating at ~28K has been observed[17]. On the other hand in polycrystalline DTO, two ferroelectric transitions at ~25K and ~13K have been observed[16]. Whereas, low temperature single crystal studies shows another ferroelectric transition at ~1.2K related to the development of spin ice configuration[18]. It was concluded that ferroelectric transitions, at ~60K in case of polycrystalline and ~28K in case of single crystal of HTO and at ~25K in polycrystalline DTO, has structural origin[15–17,19]. Whereas, low temperature ferroelectric transition around 23K (for HTO) and ~13K (for DTO) has been dedicated to originate from the magnetism of the system[15,16]. Due to cubic symmetry of space group $Fd\bar{3}m$, observed ferroelectricity is mainly originated by the breaking of inversion symmetry which locally distort the lattice. Signature of these structural distortions have been observed in Raman and principle elastic constant measurement for both compounds[19,20]. However, factor responsible for these distortions and its associated nature of ferroelectric transitions is still unknown and matter of investigation.

In this article a detailed study of the dielectric response of polycrystalline HTO and DTO has been done because of its accuracy and efficiency for the details of the local structural distortions. It is found that both HTO and DTO exhibit distinctly separable two prominent dielectric relaxations at ~90K and ~36K respectively. The observed value of critical exponent $\gamma$, which is a measure of type of phase transition obtained by modified Curie-Weiss law, confirms that it is a diffuse phase transition (DPT). Through the response of chemically induced structural distortions by Fe substitution at A & B-site, termed as HTFO and HFTO respectively, on the dielectric behavior, measured value of activation energy ($E_a$) and characteristic relaxation time ($\tau_0$), it is confirmed that both distinctly observed DPT at 90K and 36K are inherently generated by the crystal structure and related to the lattice distortions at O1 and O2 oxygen sites. Magnetic susceptibility and previously reported spin relaxation behavior confirms that crystal electric field controlled SIA behavior is responsible for these structural distortions. These observation and possible mechanism are now discussed in details.

*Crystal structure:* The crystal structure of both HTO & DTO exhibits cubic symmetry of space group Fd$\bar{3}$m, Where Ho/Dy is at 16c site and Ti at 16d site. Both sites form a three dimensional networks of corner-sharing tetrahedra which makes these lattices in the category of canonical geometrically frustrated lattices. The oxygen environment around both sites is quite interesting. Ti ions are located in octahedral voids of Fd$\bar{3}$m structure, whereas Ho/Dy ion is surrounded by eight oxygen ions as shown in fig. 1.

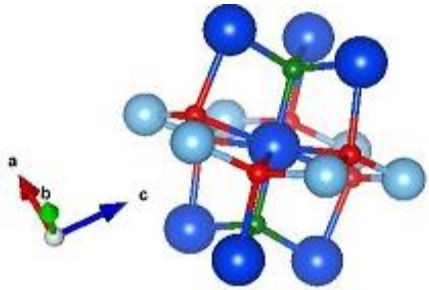

Figure 1: Oxygen environment around central Rare Earth ion (blue color). O1-site oxygen atoms (red color) are arranged in an antiprismatic manner also linked with Ti ion (sky blue color). While O2 (green color) lies perpendicular to the O1 plane.

On the basis of oxygen position around Ho/Dy ion, we can categories them in two subclasses: the O1 sites (48f-site) and O2 sites (8b-site). The O1 oxygen is displaced in two parallel planar equilateral triangles arranged in the antiprismatic manner around the central Ho/Dy ion. This antiprismatic arrangement of O1 oxygen gives the $D_{3d}$ point group symmetry which facilitates large distortion from ideal cubic structure. The remaining two oxygen ions termed as O2 are linearly aligned above and below Ho/Dy ion along local <111> axis of Fd$\bar{3}$m space group. The distance between Ho/Dy and O2 is ~2.2 Å the shortest one known for any rare earth ions, which produces a strong SIA along local <111> axis and controlled by trigonal symmetry crystal electric field (CEF)[8,10,21].

*Experimental Details:* Pure phase Polycrystalline sample HTO and DTO were synthesized by the conventional solid-state reaction methods [22]. Phase purity was confirmed by High Resolution X-ray diffraction (HRXRD) (Rigaku, Japan) having Cu K$\alpha_1$ optics near source side. Low temperature dielectric permittivity measurement was performed by using a fully computer controlled measuring system involving a Novocontrol Alpha-A high frequency analyzer and a cryogen free measurement system. Magnetic measurements were performed with SQUID Magnetic Properties Measurement System (MPMS-3)® (Quantum Design, Inc.) USA.

*Results and Discussion:* Temperature dependent dielectric measurement has been performed in a frequency range of 0.5 to 400 kHz for HTO and DTO shown in fig. 2. Two successive dielectric relaxations are observed in the temperature range 87-127K and 34-49K respectively with varying frequency, while a broad peak develops below ~30K up to lowest measured temperature in both materials. At 1 kHz, relaxations are observed at 90K and 36K in case of HTO, while in case of DTO they are observed at 92.5K and 37K.

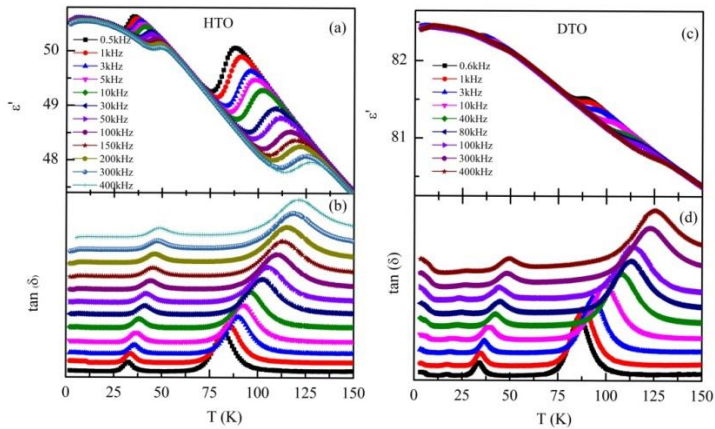

Figure 2: Temperature dependence of the dielectric permittivity and loss spectrum of HTO (a, b) and DTO (c, d) at various frequencies.

It is found that observed dielectric relaxation (~90K and ~36K) in HTO and DTO shows large frequency dispersion and the temperature ($T_m$) at which $\varepsilon'$ are maximum, shifts towards higher temperature side with increasing frequency, such a behavior representing a typical relaxation feature of relaxor ferroelectrics. However the conclusive way to define the ferroelectric phase transition is the measure of $\gamma$ in the modified Curie-Weiss law[23].

$$\frac{1}{\varepsilon'} - \frac{1}{\varepsilon_m} = \frac{(T-T_m)^\gamma}{C'} \qquad (2)$$

Where $\gamma$ and C' are constants. The critical exponent $\gamma$ is regarded as a measure of the diffuseness of the phase transition. For normal ferroelectric one gets $\gamma=1$; for relaxor phase transition one gets $\gamma=2$; while $1<\gamma<2$ correspond to the diffuse phase transition. In equation (2) the values of $\varepsilon'_m$ and $T_m$ are the maxima of permittivity and its corresponding temperature respectively. To describe the nature of phase transition in HTO as well as in DTO, we determined the critical exponent $\gamma$ using the above equation 2. Fig. 3 (a) and (b) shows the plots of $\ln(\frac{1}{\varepsilon'} - \frac{1}{\varepsilon_m})$ as a function of $\ln(T-T_m)$ for 90K and 36K relaxation at 1kHz frequency (with the calculated values of diffusivity constant, $\gamma$) for both HTO and DTO. As mentioned earlier, there are two ferroelectric transitions for both HTO and DTO. For HTO, the determined value of $\gamma$ is 1.68 and 1.26 for the phase transitions occurring at 90 K and 36 K respectively. For DTO the value of critical exponent turns out to be 1.23 for both transitions, occurring at 92.5 K and 37 K.

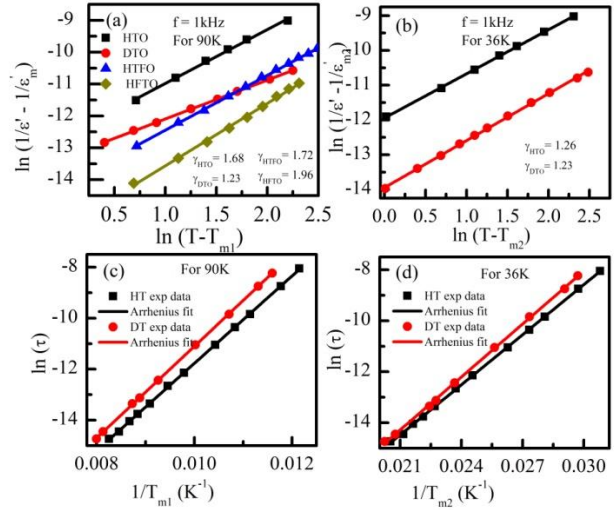

Figure 3: a & b shows the plots of ln $(1/\varepsilon' - 1/\varepsilon'_m)$ as a function of ln $(T-T_m)$ at 1 kHz (with the calculated values of diffusivity constant, γ) for 90K and 36K relaxations for HTO, DTO, HTFO & HFTO and HTO & DTO respectively. Whereas, c & d shows the plots of Arrhenius fit for 90K and 36K DPT for HTO & DTO.

The so obtained values of γ confirm that both observed dielectric relaxations are DPT for both the compounds. Diffuse or relaxor phase transitions are generally generated from the local disorder/distortion in some of the sublattices without breaking of lattice periodicity. These disorder/distortions can be created by chemical inhomogeneity e.g. cation substitution, oxygen deficiency, defects; and intrinsic crystal electric field. To investigate the possible origin of each DPT, we create chemical inhomogeneity at A-site (Ho-site) and B-site (Ti-site) through Fe-substitution. As derived from the structure, the nature of distortion is significantly different when the chemical inhomogeneities are created at these sites.

The temperature dependence of the dielectric permittivity of B site doped $Ho_2Ti_{1.85}Fe_{0.15}O_7$ (HTFO) samples is shown in figure 4(a) & (b). Out of the two ferroelectric transitions of HTO only 90 K transition is getting affected by this partial substitution. The nature of the relaxation has also changed quite significantly. The narrower transition present in case of HTO has broadened as well as the frequency dispersion smeared out over a wide temperature range. The measured value of γ (at 1 kHz) for HTFO has changed to 1.72 from 1.67 for HTO (see fig. 3a). The higher value of γ shows that the ferroelectric phase transition is shifting towards becoming a relaxor type transition but still can be categorized as DPT. This substitution at B site is most likely going to affect A/B-O1 environment only, as shown in the fig.1. It is anticipated that the DPT at 90K originate because of the distortions created at O1 site. If this hypothesis is true, then any substitution at A-site, which in turn should affect both the 90K and 36K DPTs as the substitution at A-site will affect both the A/B-O1 and A-O1/O2 environment. To verify this we have made studied the dielectric response of HTO having partial substitution at A-site as well.

The temperature dependent dielectric permittivity of $Ho_{1.8}Fe_{0.2}Ti_2O_7$ (HFTO) samples is shown in fig. 4 (c) & (d). As envisaged, A-site Fe doping affect both transition temperature and almost a single relaxation is observed in place of two relaxations in ε' and tan (δ). The measured value of γ (at 1 kHz) for HFTO turns out to be 1.96 (see fig. 3a) indicating that the ferroelectric transition becomes almost relaxor type[23,24]. The possible origin of this change in transition (diffuse to relaxor) may lie in the change in the atomic environment of both sites as discussed above.

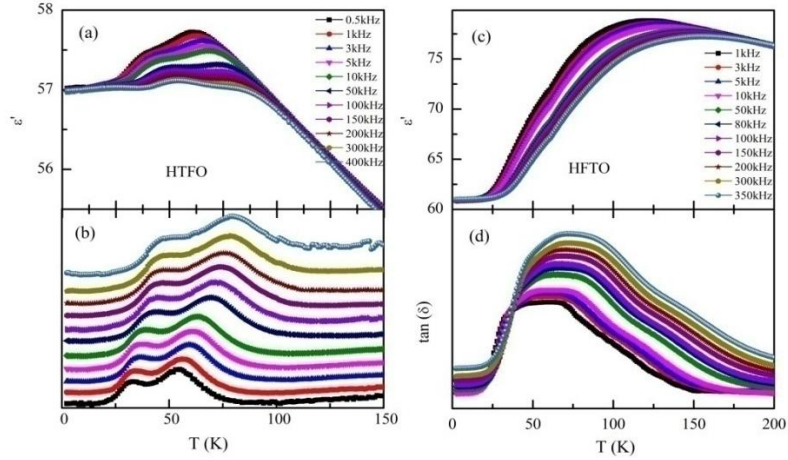

Figure 4: Temperature dependence of the dielectric permittivity of HTFO (a) and HFTO (b) at various frequencies.

These results indicate that out of the two diffused ferroelectric phase transitions prevailing in HTO at 90K and 36K is originating from O1 site distortions and O1/O2 site distortions respectively. Further, A-site Fe substitution, due to smaller ionic size than Ho, largely distort both the local A-O1 and A-O2 bond length and bond angle, resulting a greater disorder and thus changing the phase transition from diffuse to relaxor type. A similar observation has been reported by Lin et al. where A-site Gd and Tb substituted DTO confirmed that 25 K ferroelectric transition is related to A-site Dy ion[16].

Subsequently, to investigate the possible cause for these distortions, we analyze the frequency dispersion behaviour using Arrhenius law:

$$\tau = \tau_0 \exp\left(\frac{E_a}{KT}\right) \quad (3)$$

Where $\tau$ is the measuring relaxation time in sec, $\tau_0$ is the characteristic relaxation time, and $E_a$ is the activation energy given in the unit of meV. Fig. 3c & d shows a plot of relaxation time ($\tau$) vs.

$T_m$, of both 90 and 36 K DPTs for both HTO and DTO compounds. The calculated value of $E_a$ and $\tau_0$ for HTO and DTO are summarized in the table given below.

Table I: value of activation energy $E_a$ (meV) and characteristic relaxation time $\tau_0$ (sec) for 90K and 36K DPT as obtained from Arrhenius fit.

| Sample | $T_{m1}$ | | $T_{m2}$ | |
|---|---|---|---|---|
| | $E_a$ (meV) | $\tau_0$ (sec) | $E_a$ (meV) | $\tau_0$ (sec) |
| HTO | 147.6±0.5 | 2.7× 10$^{-13}$ | 56.4±0.26 | 5.5× 10$^{-13}$ |
| DTO | 154.8±0.5 | 2.3× 10$^{-13}$ | 59.6±0.4 | 3.0× 10$^{-13}$ |

The observed value of characteristic relaxation time for both the relaxations has same order but different value of activation energies. Relaxation time originating from polar lattice vibrational effects lies in the range of $10^{-9}$-$10^{-14}$ sec. In the case of HTO and DTO has the observed value of the $\tau_0$ is in range of $2 - 6 \times 10^{-13}$ sec. This indicated that that observed DPT in case of HTO and DTO are related to polar lattice vibrational effects originating from the local distortions. The order of characteristic relaxation time and activation energy as obtained by Arrhenius fit are well correlated with other relaxor ferroelectric e.g. $Bi_{1.5}Zn_{1.0}Nb_{1.5}O_7$ and Bi2O3–ZnO–Nb2O5 cubic pyrochlores[25,26].

In the present studies of HTO and DTO, the observed value of activation energy (~50meV) for 36K DPT is relatively small when compared to the same of chemically doped pyrochlores. The bond length of A-O2 is one of the smallest bond lengths between Rare Earth and oxygen in the pyrochlore structure and most susceptible for any distortion. It is highly probable that the observed 36K DPT is originated by local structural distortion in O2 site oxygen position and influenced by the phonon branches associated with O2-A-O2 bending modes[25,26]. Whereas 90K

DPT is originated by local structural distortion in O1 site oxygen position and must be influenced by the phonon branches associated with O1-A-O1 bending modes and A-BO6 stretching modes according to their order of activation energy (~150meV).

In pyrochlores, ferroelectric and multipolar phases should have an electronic origin due to the incompatibility of the gyrotropic order (which belongs to the $A_{1u}$ representation of the $O_h$ point group) with any phonon mode at the Brillouin zone center[27] and breaks the parity of the crystal. The parity breaking order parameter is time-reversal invariant and break the rotational symmetry of the crystal, which linearly coupled with distortions in the lattice sites[28] as observed in the form of DPTs in the dielectric response in this case. In case of HTO and DTO spin ices, Ho/Dy ions have large magnetic moment and govern by crystal electric field which is likely to affect A-site crystal environment. This kind of interplay of magnetic anisotropy with local structural distortion is well known. DFT calculations by Weingart et al. shows that the exact type of magnetic anisotropy crucially dependent on the details of the local distortions of the perovskite structure[29]. In neutron scattering studies of HTO, an abrupt increase in spin relaxation time has been observed at ~80K which becomes saturated at ~30K. Whereas, below 30K, a temperature independent plateau was observed up to ~3K[30]. A plot of temperature dependence of the inverse of magnetic susceptibility and dielectric permittivity at 100 kHz for HTO is shown in fig. 5. A straight line is fitted (solid line) in the temperature range 300-200K and extrapolated up to 2K. A clear deviation from Curie-Weiss behaviors at ~130K in inverse of magnetic susceptibility started and becomes prominent as temperature decreases. It should be noted that the first DPT is observed in this temperature range ~87K–127K only for measured frequency range.

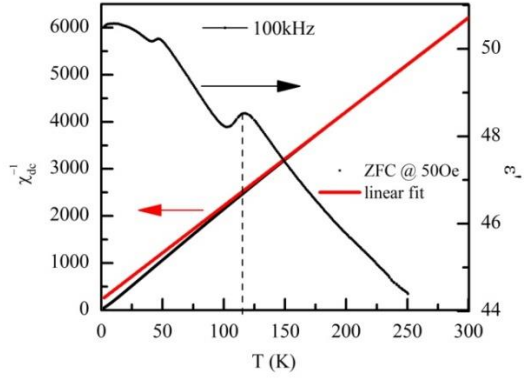

Figure 5: Temperature dependence of inverse magnetic susceptibility (left axis) and dielectric permittivity at 100 kHz (right axis) for HTO. The solid line represents the line fit of magnetic susceptibility in temperature range 300-200K and extrapolated up to 2K.

These observations clearly indicate that observed DPTs and spin relaxation behavior are SIA induced and interconnected with each other. It is suggested that SIA coefficients have thermal variation similar to other RE compounds[31,32]. This thermal variation of SIA coefficient could be responsible for an abrupt spin orientation of RE moment towards the local <111> axis. At higher temperature (above ~130K) RE spins are most likely non-Ising type, creates an isotropic field around rare earth ion. However, with decreasing temperature the CEF controlled SIA, vector component of magnetic moment gains its magnitude and directionality along local <111> axis[8,11], with a simultaneous increase in spin relaxation time. This process gets completed when Ho/Dy moment becoming parallel to the <111> axis. These spin orientation process (non-Ising to Ising) is mainly occur in the temperature range 80-30K[30]. In neutron scattering studies it has been confirmed for dipolar spin ice $Ho_2Sn_2O_7$[33] as well, that Ising nature of spin existed only below 40K. Non-Ising to Ising spin crossover trigger a huge change in the mean field around each RE-site, thus changing the inter-ionic electrostatic coulomb interactions, in case of both HTO and DTO, with the decrease in temperature. Since planer O1-site is perpendicular to the local <111> axis due to which a large decrease in mean field takes place in this plane. To screen this change, a small displacement/distortion in planer O1 site oxygen atom is likely to takes place via the bond bending in O1-A-O1 which gets facilitated by A-site $D_{3d}$ point group symmetry. This bond bending also distort the B-O6 octahedron through O1-B-O1 bending, B-O1 and O1-O1 stretching, which facilitate a large increment in the trigonal CEF acting on Ho/Dy ion[19,34], resulting an abrupt increase in SIA coefficient.

Similar phenomena may takes place for 36K DPT, as component of the magnetic moment of RE along local <111> axis i.e. towards O2 site increases with decreasing temperature, a distortion at this site take place due to increase in mean field influenced by the phonon branches associated with O2-A-O2 bond bending[25,26]. Since O2 oxygen atom lies at the centre of each corner sharing tetrahedra formed by A-site RE ion, a huge electrostatic mean field could get created at O2 site by Ho/Dy. These Ising spins are fluctuating in different microstates (which are 16) with equal probability thus producing a mean electrostatic field on each central O2-site oxygen atom. This mean field constrained thermal motion of O2 site oxygen along their <111> axis, resulting a displacement along other direction takes place. In these directions, oxygen atoms are hopping independently with equal magnitude but larger than polar lattice vibration, results into a well defined ferroelectric diffused phase transition at 36K governed by O2-site.

*Conclusion:* In short, both observed dielectric relaxation at ~90K and ~36K is a diffuse ferroelectric phase transition and distinctly related to the lattice distortions at the O1 (48f) and O2 (8b) crystallographic sites of oxygen. In these compounds, due to the incompatibility of the gyrotropic order with any phonon mode at the Brillouin zone center, observed diffuse ferroelectric phase transitions can have only an electronic origin. Based on magnetic susceptibly and previously reported spin relaxation behavior, it has been concluded that temperature dependent single ion anisotropy behavior and it's coupled spin orientation of the rare earth magnetic moment from isotropic non-Ising to Ising towards their local <111> axis break rotational symmetry of the lattice. Due to the linear coupling of rotational symmetry with lattice, it distorts the O1 and O2 oxygen sites distinctly as observed in the form of 90K and 36K DPTs. This provides a possible explanation of the observed multiferroic nature in these materials, which might be helpful for designing the other geometrically frustrated multiferroic materials.